\documentclass[twocolumn,showpacs,preprintnumbers,amsmath,amssymb]{revtex4-1}

\usepackage{afterpage}
\usepackage[dvipdfmx]{graphicx}
\usepackage{color,bm}

\begin{document}

\title{
Gravitational lens without asymptotic flatness: 
Its application to the Weyl gravity
} 
\author{Keita Takizawa}
\author{Toshiaki Ono}
\author{Hideki Asada} 
\affiliation{
Graduate School of Science and Technology, Hirosaki University,
Aomori 036-8561, Japan} 
\date{\today}

\begin{abstract} 
We discuss, without assuming asymptotic flatness, 
a gravitational lens for an observer and source that 
are within a finite distance from a lens object. 
The proposed lens equation is consistent with the deflection angle of light 
that is defined for the nonasymptotic observer and source by Takizawa et al. 
[Phys. Rev. D 101, 104032 (2020)] 
based on the Gauss-Bonnet theorem with using the optical metric. 
This lens equation, 
though it is shown to be equivalent to 
the Bozza 
lens equation [Phys. Rev. D 78, 103005 (2008)],  
is linear in the deflection angle. 
Therefore, the proposed equation is more convenient 
for the purpose of doing an iterative analysis. 
As an explicit example of an asymptotically nonflat spacetime, 
we consider a static and spherically symmetric solution 
in Weyl conformal gravity, 
especially a case that $\gamma$ parameter in the Weyl gravity model 
is of the order of the inverse of the present Hubble radius. 
For this case, we examine iterative solutions  
for the finite-distance lens equation up to the third order. 
The effect of the Weyl gravity on the lensed image position 
begins at the third order and it is linear in the impact parameter of light. 
The deviation of the lensed image position from the general relativistic one 
is $\sim 10^{-2}$ microarcseconds for the lens and source with a separation angle of 
$\sim 1$ arcminute, 
where we consider a cluster of galaxies 
with $10^{14} M_{\odot}$ at $\sim 1$ Gpc for instance. 
The deviation becomes $\sim 10^{-1}$ microarcseconds, 
even if the separation angle is $\sim 10$ arcminutes. 
Therefore, effects of the Weyl gravity model 
are negligible in current and near-future observations of gravitational lensing. 
On the other hand, the general relativistic corrections at the third order 
$\sim 0.1$ milliarcseconds 
can be relevant with VLBI observations. 
\end{abstract}


\maketitle

\section{Introduction}
The gravitational deflection of light has been an important tool 
in gravitational physics, since it was measured by Eddington and his collaborators 
\cite{Eddington}.  
Gravitational lens has been one of the key subjects 
in the modern astronomy and cosmology, 
though Einstein thought that the phenomenon of a star 
acting as a gravitational lens was unobservable 
\cite{Einstein1936}. 
In particular, the Event Horizon Telescope (EHT) team has recently succeeded 
a direct imaging of the immediate vicinity of the central black hole candidate 
of M87 galaxy \cite{EHT}. 

The formulation of the gravitational lens and its applications are usually 
based on the gravitational lens equation. 
The conventional lens equation uses the deflection angle of light 
that is defined for the asymptotic receiver (denoted by R) 
and source (denoted by S), 
where the observer is referred to the receiver 
in order to avoid a confusion in notations 
between $r_0$ (the closest approach of light) 
and $r_O$ by using $r_R$. 

Gibbons and Werner proposed an alternative way of defining 
the asymptotic deflection angle of light \cite{GW}, 
where the receiver and source of light are assumed to be 
in an asymptotically Minkowskian region. 
The Gauss-Bonnet theorem \cite{GBMath} 
with using the optical metric plays a crucial role 
in their geometrical definition of the deflection angle. 
Their method has been vastly applied to a lot of spacetime models 
especially by Jusufi and his collaborators 
e.g. \cite{Jusufi2017a, Jusufi2017b, Jusufi2018},  
and has been extended to study the gravitational deflection of light 
in a plasma medium 
e.g. \cite{Crisnejo2018, Crisnejo2019}. 

Ishihara et al. extended the idea of Gibbons and Werner 
to study effects of finite distance on the gravitational deflection of light, 
where the receiver and source are within a finite distance from a lens object 
\cite{Ishihara2016, Ishihara2017}. 
Their formulation has been extended to stationary and axisymmetric 
spacetimes such as Kerr solution \cite{Ono2017}, 
a rotating wormhole \cite{Ono2018} and 
a rotating global monopole with an angle deficit \cite{Ono2019}. 
Their definition of the deflection angle is still limited within 
asymptotically flat spacetimes. 
See Reference \cite{Ono2019b} for a review on this subject. 

Without assuming asymptotic flatness, Takizawa et al. proposed 
a definition of the gravitational deflection of light 
for the receiver and source that are within a finite distance from a lens object 
\cite{Takizawa2020}. 
In their definition based on the Gauss-Bonnet theorem, 
the radial interval is exactly the same as that for the light ray from the source 
to the receiver. 
As a result, this definition can be applied not only to an asymptotically flat 
black hole but also to an asymptotically nonflat black hole such as 
the Kottler (Schwarzschild-de Sitter) solution in general relativity and 
a static and spherically symmetric vacuum solution in Weyl conformal gravity. 

The deflection angle of light is not always observable. 
As mentioned above, the gravitational lensing observables are 
discussed by using the gravitational lens equation. 
How can the deflection angle of light for nonasymptotic receiver and source 
be incorporated into the gravitational lens equation? 
The main purpose of this paper is 
to discuss a gravitational lens equation valid for the deflection angle of light 
that is defined by Takizawa et al. \cite{Takizawa2020}, 
without assuming asymptotic flatness,  
for an observer and source within a finite distance from a lens object.

This paper is organized as follows. 
In Section II, the lens equation with finite-distance effects is reexamined. 
In Section III, 
we discuss iterative solutions for the finite-distance lens equation 
in the small angle approximation. 
Section IV discusses the lensed image positions 
in a static, spherically symmetric vacuum solution in Weyl conformal gravity. 
In Section V, we examine whether effects of Weyl conformal gravity 
on the gravitational lens can be tested by present and near-future astronomical 
observations. 
Section VI is devoted to the conclusion. 
Throughout this paper, we use the unit of $G = c = 1$.

\section{Lens Equation in a finite-distance situation}
\subsection{Effect of finite distances on the light propagation}
We follow References \cite{Ishihara2016, Takizawa2020} to 
consider a static and spherically symmetric spacetime. 
The metric reads 
\begin{align}
ds^2 &= g_{\mu\nu} dx^{\mu} dx^{\nu} 
\nonumber\\
&= -A(r) dt^2 + B(r) dr^2 + C(r) d\Omega^2 , 
\label{ds2-SSS}
\end{align}
where 
$d\Omega^2 \equiv d\theta^2 + \sin^2\theta d\phi^2$ and 
$\phi$ is the azimuthal angle respecting the rotational symmetry.  
If we choose $C(r) = r^2$, then, $r$ denotes the circumference radius.  
Henceforth, we choose the photon orbital plane as 
the equatorial plane without the loss of generality, 
because the spacetime is spherically symmetric.

\begin{figure}
\includegraphics[width=8.6cm]{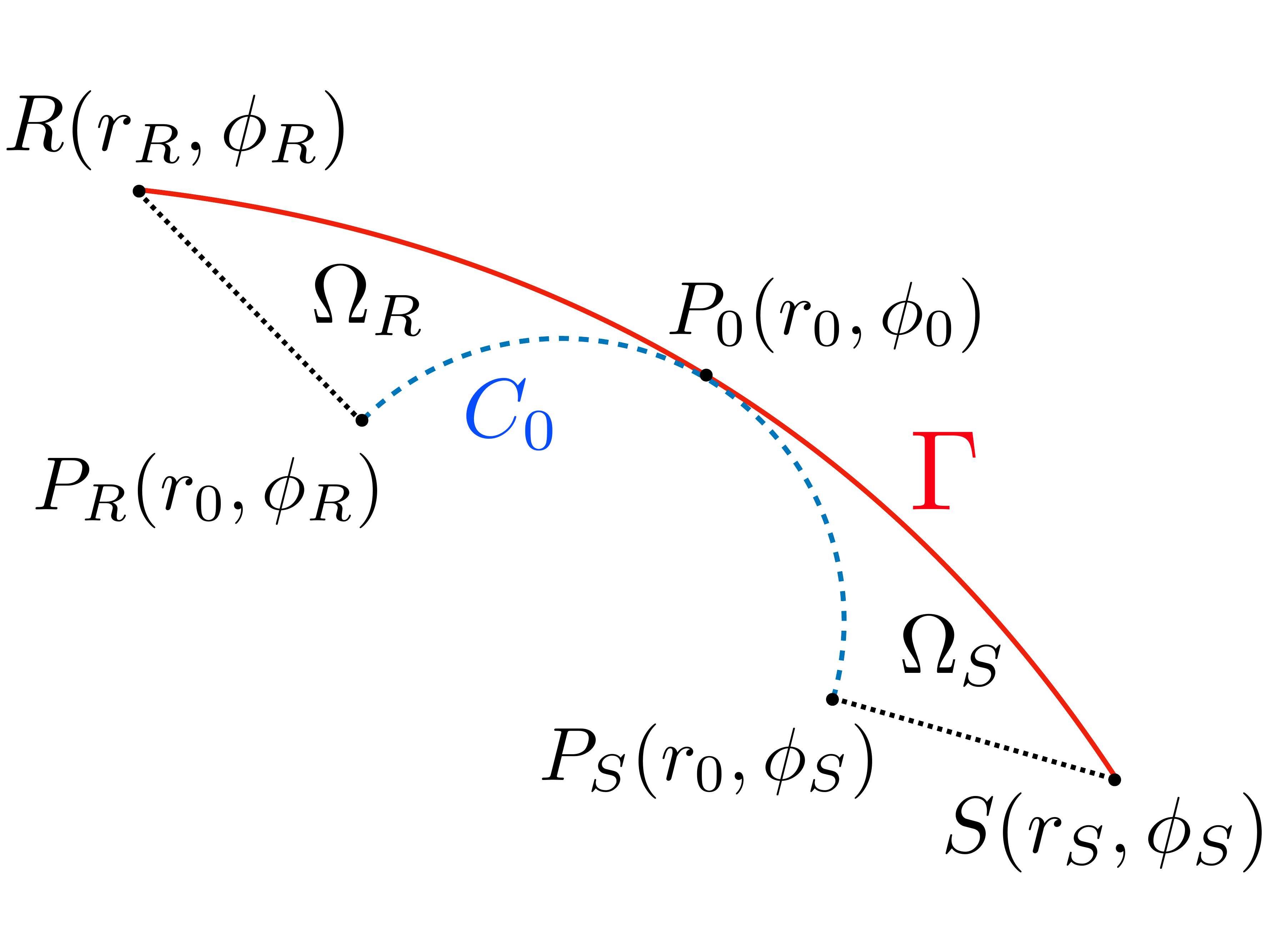}
\caption{
$\Omega_R$ and $\Omega_S$. 
$\Omega_R$ is a trilateral specified by the points $R$, $P_0$ and $P_R$. 
$\Omega_S$ is that specified by the points $S$, $P_0$ and $P_S$. 
}
\label{fig-Kitazawa}
\end{figure}

In order to avoid requiring the asymptotic flatness of a spacetime, 
Takizawa et al. proposed an integral form of the definition 
for the deflection angle of light (denoted as $\alpha_K$) 
for an observer and source that 
are within a finite distance from a lens object \cite{Takizawa2020}. 
$\alpha_K$ is defined as 
\begin{align}
\alpha_K 
\equiv 
\iint _{\Omega_R + \Omega_S} K dS + \int_{P_R}^{P_S} \kappa_g d\ell + \phi_{RS} . 
\label{alpha-K}
\end{align}
The right-hand side of this equation 
contains the radial coordinate $r \in [r_0, r_R]$ or $[r_0, r_S]$, 
where $r_0$ means the closest approach of light. 
Indeed, this radial interval is exactly the same as that for 
the light ray from the source to the receiver. 
See Figure \ref{fig-Kitazawa}.

\begin{figure}
\includegraphics[width=8.6cm]{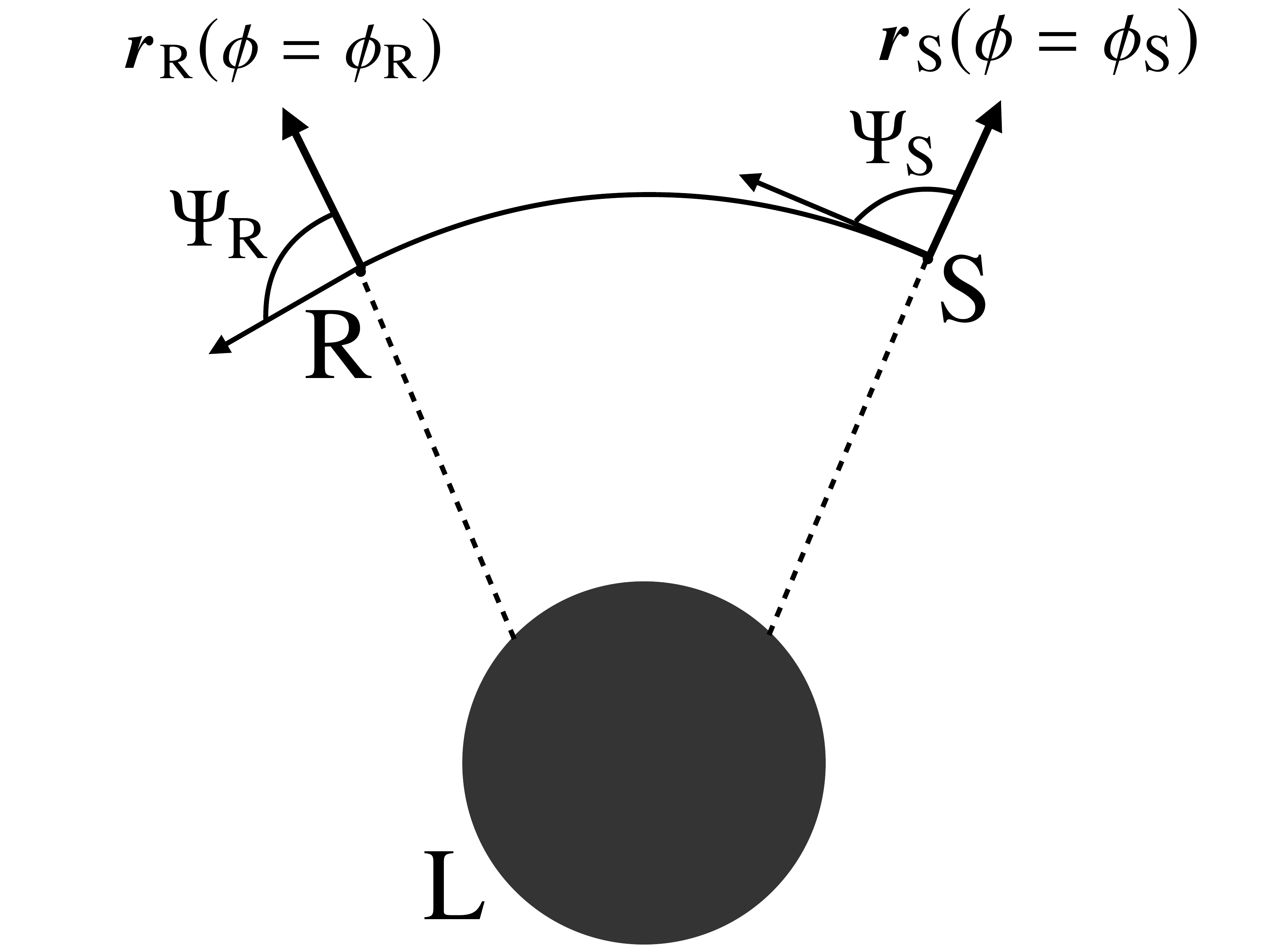}
\caption{
The light ray and radial directions. 
The angle between the light ray and the radial direction at the receiver 
is $\Psi_R$ and 
that at the source is $\Psi_S$. 
The coordinate angle between the receiver and the source 
is $\phi_{RS} = \phi_R - \phi_S$. 
}
\label{fig-Ishihara}
\end{figure}

Without assuming the asymptotic flatness, 
they proved that 
their definition agrees with another form of the deflection angle 
by Ishihara et al. \cite{Ishihara2016} 
which assumed the asymptotic flatness.  
Ishihara et al. \cite{Ishihara2016} defined the deflection angle of light as 
\begin{align}
\alpha_I \equiv \Psi_R - \Psi_S + \phi_{RS} , 
\label{alphaI}
\end{align}
where $\Psi_S$ and $\Psi_R$ are the angles between the radial direction 
and the light ray at the source position and at the receiver position, respectively, 
and $\phi_{RS}$ is a coordinate angle between the receiver and source. 
See Figure \ref{fig-Ishihara} for these angles.  

It was shown that 
\begin{align}
\alpha_I = \alpha_K ,  
\label{alphaIK}
\end{align}
holds in general for a static and spherically symmetric spacetime, 
especially even for an asymptotically nonflat case 
\cite{Takizawa2020}.

\subsection{Finite-distance expressions for the deflection angle of light}
\begin{figure}
\includegraphics[width=8.6cm]{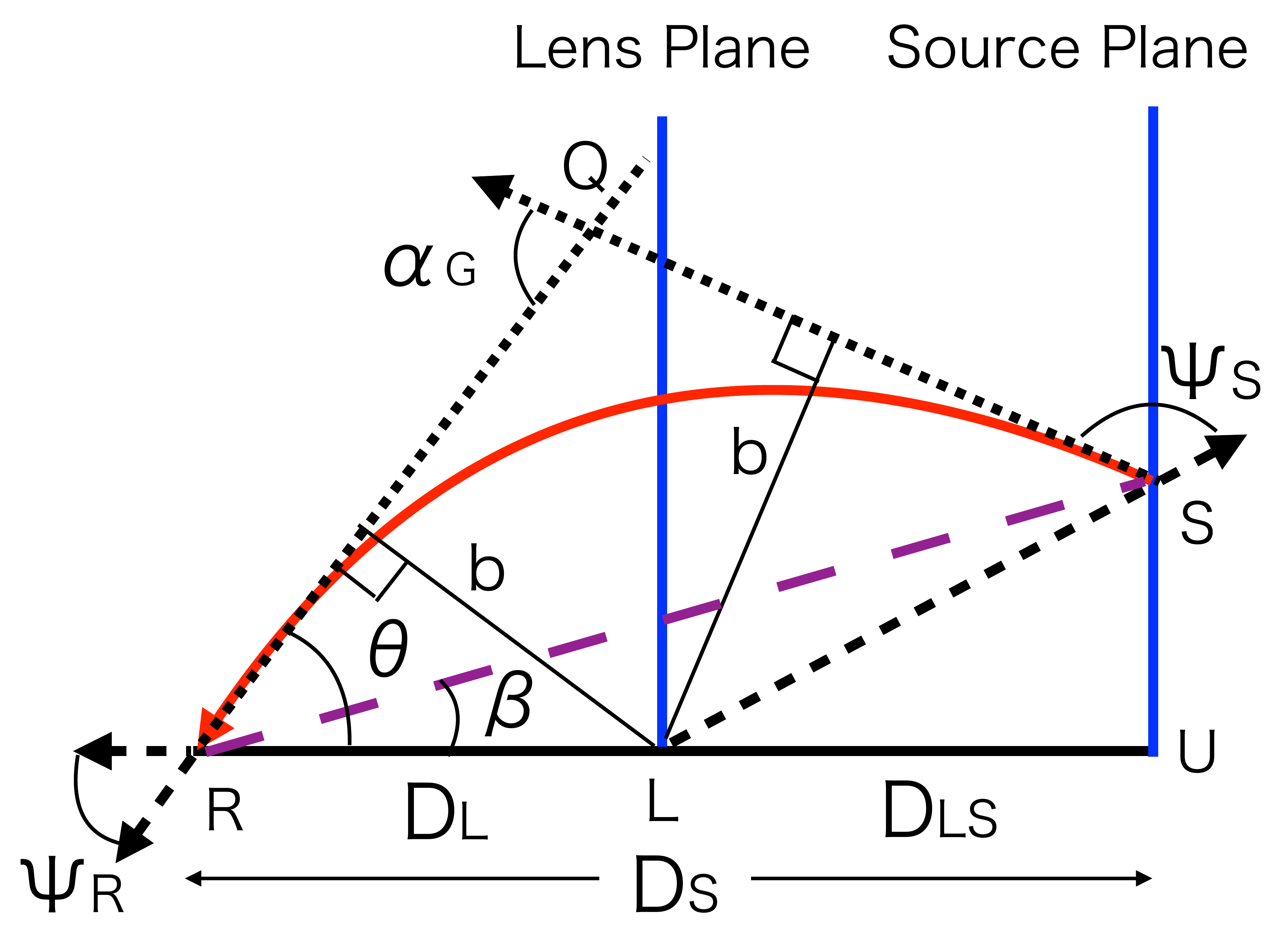}
\caption{
Geometrical gravitational lensing setup. 
}
\label{fig-setup}
\end{figure}

We introduce the lens plane and the source one 
to examine the gravitational lens equation. 
See Figure \ref{fig-setup} for the gravitational lensing configuration 
in this paper, where the thin lens approximation is not used. 
The (red in color) solid curve in this figure shows 
the light ray from the source to the receiver. 
The angles $\Psi_R$ and $\Psi_S$ appear in Eq. (\ref{alphaI}). 
The tangents at the receiver and the source are denoted 
by the dotted lines in this figure. 
These tangent lines intersect at the point Q. 
Note that the intersection point Q is not necessarily in the lens plane. 
In the conventional formulation with the thin lens approximation 
for the asymptotic receiver and source, the intersection point 
is often assumed implicitly to be on the lens plane. 
The assumption that the intersection point is in the lens plane 
needs a symmetric configuration in which 
the receiver and source are equidistant from the lens. 
This additional assumption is made also in Virbhadra and Ellis 
for their formulation of the almost exact lens equation, 
though this formulation is valid not only for the weak deflection 
but also for the strong deflection \cite{VE2000}. 
See e.g. Figure 1 and the paragraph including Eqs. (1)-(3) 
in Reference \cite{VE2000}. 
From the aspect of the triangular inequality, 
Dabrowski and Schunck realized 
difficulties of using Virbhadra and Ellis lens equation and derived 
an alternative  lens equation 
\cite{DS}. 
However, Dabrowski and Schunck lens equation still relies upon  
the additional assumption that the intersection point Q 
lies on the lens plane. 
See Eq. (23) and Appendix in Reference \cite{DS}. 
The additional assumption of the intersection point lying on the lens plane 
was argued also by Bozza \cite{Bozza2008}.

$D_L$, $D_S$ and $D_{LS}$ denote 
the angular diameter distances from the receiver to the lens, 
from the receiver to the source and from the lens to the source, respectively. 
The angular direction of the lensed image with respect to the lens direction 
is denoted by $\theta$ 
and that of the intrinsic source position is denoted by $\beta$. 
These angles $\theta$ and $\beta$ are defined at the receiver point. 
$\theta$ equals to $\Psi_R$. 
See also Figure \ref{fig-setup}.

\begin{figure}
\includegraphics[width=8.6cm]{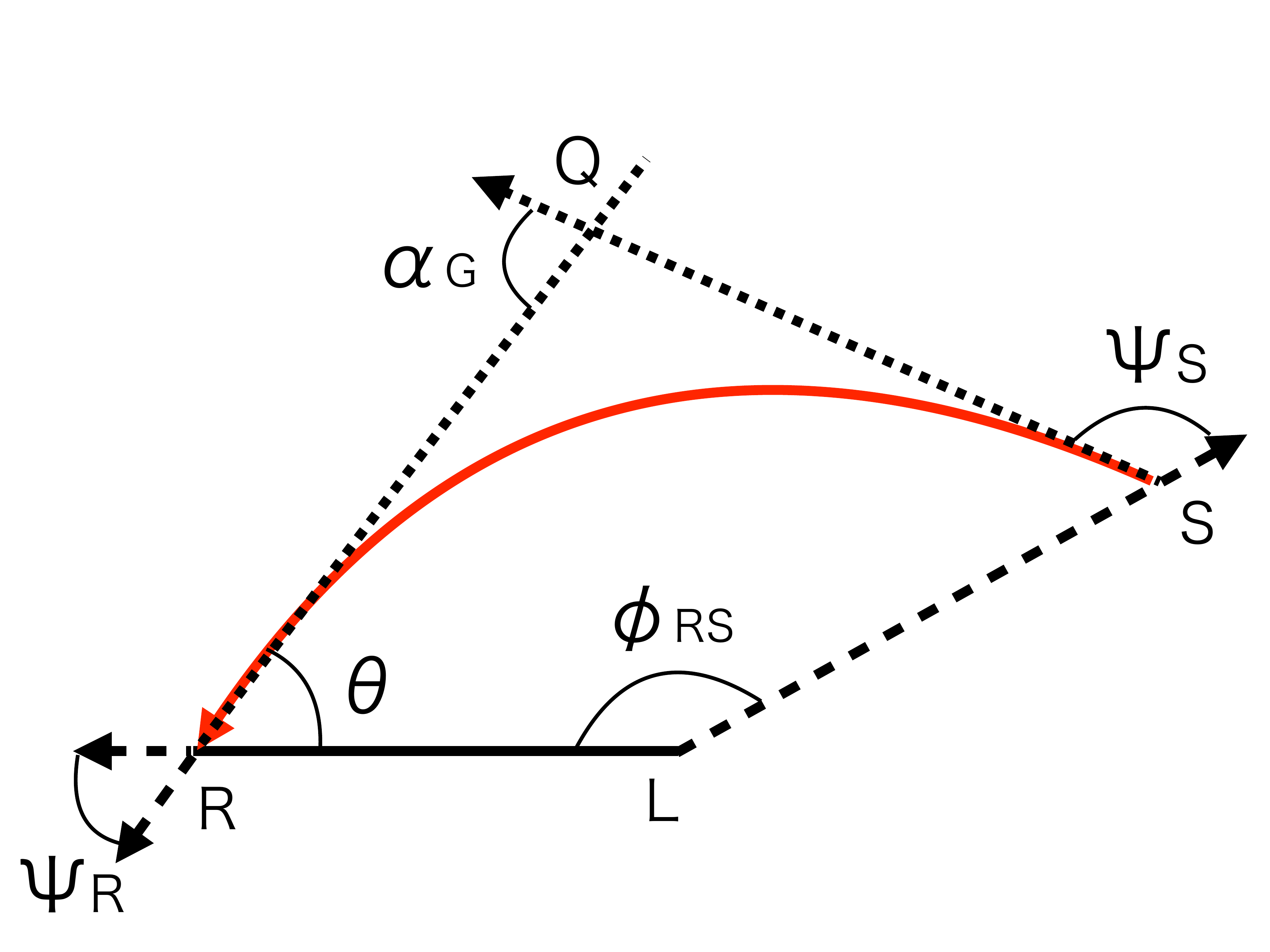}
\caption{
Quadrilateral LRQS in the geometrical gravitational lensing configuration. 
This is corresponding to Figure \ref{fig-setup}. 
The directional difference between the receiver and source is assumed to be 
$\phi_{RS}$ at the point L. 
The (red in color) solid curve denotes the light ray from the source to the receiver. 
The angle between the two tangent lines  in this figure 
is interpreted as the deflection angle of light. 
The deflection angle is denoted as $\alpha_G$. 
}
\label{fig-quadrilateral}
\end{figure}

We consider a quadrilateral LRQS in Figure \ref{fig-setup}. 
Figure \ref{fig-quadrilateral} focuses on the quadrilateral LRQS. 
In this geometrical configuration of the gravitational lensing, 
we define the deflection angle of light $\alpha_G$ as 
the angle at the point Q between these tangent lines. 
In the gravitational lensing interpretation, 
the inner angle at the lens in LRQS is assumed to be $\phi_{RS}$. 
For the quadrilateral, we obtain 
\begin{align}
\theta + \phi_{RS} + (\pi - \Psi_S) + (\pi - \alpha_G) = 2 \pi , 
\label{quadrilateral}
\end{align}
where we follow the gravitational lensing interpretation to assume 
that the sum of the inner angles in any convex quadrilateral is $2\pi$. 
By using Eq. (\ref{quadrilateral}), 
we define $\alpha_G$ as 
\begin{align} 
\alpha_G \equiv \theta - \Psi_S + \phi_{RS} . 
\label{alphaG}
\end{align}

From Eqs. (\ref{alphaI}) and (\ref{alphaG}), we find 
\begin{align}
\alpha_I = \alpha_G , 
\label{alphaGI}
\end{align}
where we use $\Psi_R = \theta$. 
Therefore, $\alpha_I$ defined by Eq. (\ref{alphaI}) 
can be safely interpreted as the deflection angle of light.

From Eqs. (\ref{alphaIK}) and (\ref{alphaGI}), we obtain the equivalence of 
the three definitions of the deflection angle of light, namely 
\begin{align}
\alpha_G = \alpha_I = \alpha_K .
\label{alphaGIK}
\end{align}
In the following, we use $\alpha_G$ to study the gravitational lens equation, 
because $\alpha_G$ is written in terms of $\theta$ that plays a crucial role 
in the gravitational lens equation.

\begin{figure}
\includegraphics[width=8.6cm]{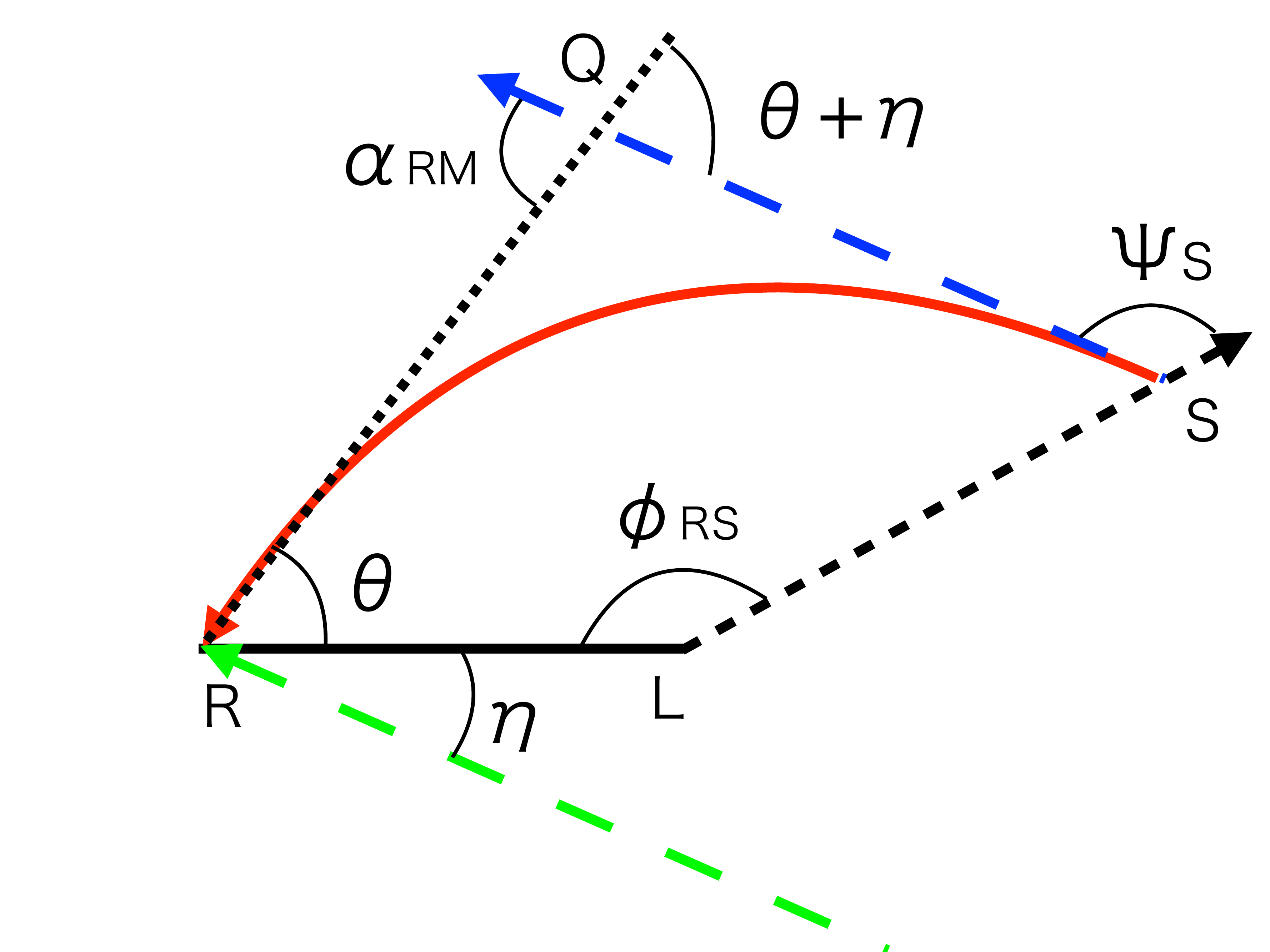}
\caption{
Gravitational lensing setup in Richter and Matzner method 
for the PPN metric \cite{RM}. 
As a reference, they assume a (green in color) dashed line from the receiver. 
Here, the (green in color) dashed line is supposed to be parallel 
to the (blue in color) dashed line 
that is tangent to the light ray at the source. 
The deflection angle can be defined as 
$\alpha_{RM} \equiv \theta + \eta$, 
though the sign convention for $\eta$ in this figure 
is opposite to that by Richter and Matzner
}
\label{fig-RM}
\end{figure}

Before going to detailed calculations of $\alpha_G$, 
we briefly mention another finite-distance expression 
of the deflection angle (denoted as $\alpha_{RM}$) 
computed by Richter and Matzner for the PPN metric \cite{RM}. 
The equivalence between $\alpha_I (=\alpha_G)$ and $\alpha_{RM}$  
was noticed by Crisnejo et al. \cite{Crisnejo2019b}. 
See Figure \ref{fig-RM} for the lensing setup in $\alpha_{RM}$. 
Figures \ref{fig-quadrilateral} and \ref{fig-RM} show 
$\alpha_G = \alpha_{RM}$. 
It is worthwhile to point out 
that the definition of $\alpha_{RM}$ needs 
a comparison between the two parallel lines (in Figure \ref{fig-RM}) 
and hence $\alpha_{RM}$ is rather limited compared with $\alpha_G$.

\subsection{Effect of finite distances on the gravitational lens equation} 
$\alpha_G$ is a relation among angles, in which any distance 
does not explicitly appear. 
Therefore, we shall study some relations between angles and distances. 
The light ray (red solid curve in Figure \ref{fig-setup}) 
is specified by the impact parameter of light (denoted as $b$). 
At the point R, this is described by 
\begin{align} 
b = LR \sin \theta .
\label{b-theta}
\end{align}
At the point S, it is expressed as 
\begin{align}
b = LS \sin (\pi -\Psi_S) .
\label{b-PsiS}
\end{align}

The impact parameter $b$ is common to Eqs. (\ref{b-theta}) and (\ref{b-PsiS}), 
so that $b$ can be eliminated as 
\begin{align}
LS \sin (\pi -\Psi_S) = LR \sin \theta .
\end{align} 
This is solved for $\Psi_S$ as 
\begin{align}
\Psi_S = \pi - \arcsin\left(\frac{LR}{LS} \sin \theta \right) .
\label{PsiS}
\end{align} 

We consider the triangles RSU and LSU in Figure \ref{fig-setup}. 
The length SU is written in two ways as 
\begin{align}
SU &= LS \sin (\pi - \phi_{RS}) , 
\\
SU & = RS \sin \beta . 
\end{align}
By eliminating SU from these equations, 
\begin{align}
LS \sin (\pi - \phi_{RS}) = RS \sin \beta . 
\end{align}
Hence, we obtain 
\begin{align}
\phi_{RS} = \pi - \arcsin \left( \frac{RS}{LS} \sin \beta \right). 
\label{phiRS}
\end{align}

Substituting Eqs. (\ref{PsiS}) and (\ref{phiRS}) into Eq. (\ref{alphaG}), 
we obtain 
\begin{align}
\alpha_G - \theta 
= \arcsin\left(\frac{LR}{LS} \sin \theta \right) 
- \arcsin \left( \frac{RS}{LS} \sin \beta \right) . 
\label{lenseq0}
\end{align}
$LR = D_L$ is a constant in the gravitational lensing formulation. 
On the other hand, $LS$ and $RS$ are dependent on 
the source position described by the parameter $\beta$. 
We thus rewrite them in terms of the angular diameter distances 
$D_L$, $D_S$ and $D_{LS}$.

\begin{figure}
\includegraphics[width=8.6cm]{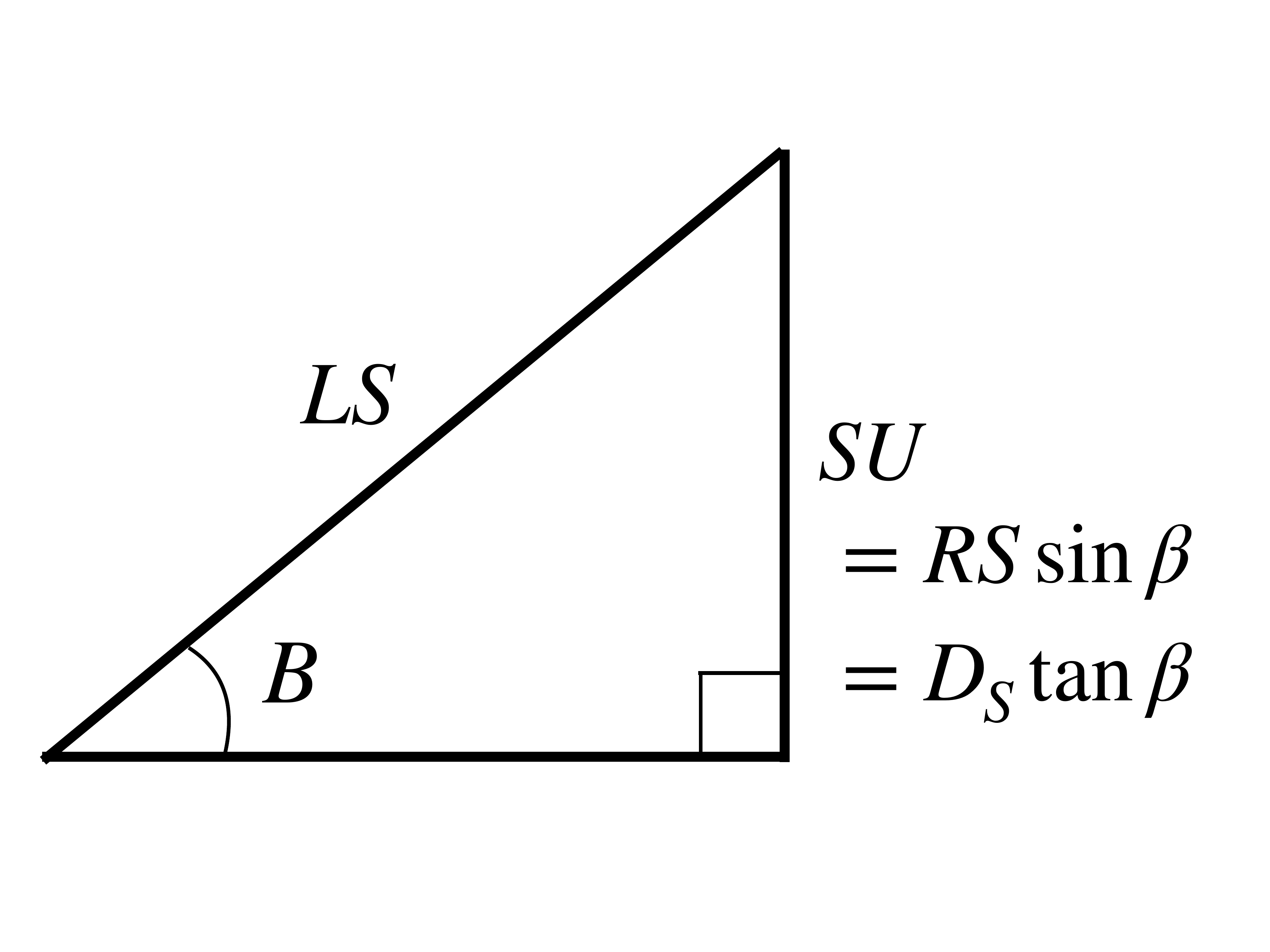}
\caption{
Geometrical meaning of $B$. 
The angle $B$ is defined by Eq. (\ref{B}). 
}
\label{fig-B}
\end{figure}

Let $B$ denote the second term in the right-hand side of 
Eq. (\ref{lenseq0}). 
Namely, it is defined by 
\begin{align}
B \equiv \arcsin \left( \frac{RS}{LS} \sin\beta \right) .
\label{B}
\end{align}
The term $B$ means an angle in a triangle by Figure \ref{fig-B}. 
Note that this triangle does not appear in the lensing configuration 
by Figure \ref{fig-setup} as it is. 
The length of the base for this triangle in Figure \ref{fig-B} becomes 
\begin{align}
\sqrt{(LS)^2 - (RS \sin\beta)^2} = D_{LS} . 
\label{base}
\end{align} 
Here, we used 
\begin{align}
SU &= RS \sin\beta 
\notag\\
&= D_S \tan\beta , 
\label{SU}
\end{align} 
and 
\begin{align}
(LS)^2 - (SU)^2 = (D_{LS})^2 . 
\label{Pythagorean} 
\end{align} 
The first relation is obtained by using Figure \ref{fig-setup}  
and the second one can be derived from the triangle LSU. 

Eq. (\ref{base}) is used for the triangle in Figure \ref{fig-B}. 
B is thus rewritten in terms of the angular distances as  
\begin{align}
B = \arctan \left( \frac{D_S}{D_{LS}} \tan\beta \right) . 
\label{B2} 
\end{align}
In the similar manner, we use Eqs. (\ref{SU}) and (\ref{Pythagorean}) to obtain 
\begin{align}
LS 
&= \sqrt{(D_{LS})^2 + (SU)^2} 
\notag\\
&= \sqrt{(D_{LS})^2 + (D_S)^2 \tan^2 \beta} .  
\label{LS} 
\end{align}

Eqs. (\ref{B2}) and (\ref{LS}) are substituted into the second and first terms 
in the right-hand side of Eq. (\ref{lenseq0}), respectively. 
We thus obtain 
\begin{align}
&\alpha_G - \theta 
- \arcsin\left(\frac{D_L}{\sqrt{(D_{LS})^2 + (D_S)^2 \tan^2 \beta}} \sin \theta \right) 
\notag\\
&+ \arctan \left( \frac{D_S}{D_{LS}} \tan\beta \right)  
\notag\\
&= 0 , 
\label{lenseq} 
\end{align}
where we used $LR = D_L$. 
Eq. (\ref{lenseq}) is the gravitational lens equation, 
in the sense that it is an equation for the lensed image position $\theta$ 
when the intrinsic source position $\beta$ and the angular distances $D_L$, $D_S$ 
and $D_{LS}$ are given. 
We should stress that Eq. (\ref{lenseq}) is linear in $\alpha_G$. 
This linearity makes perturbative calculations much simpler 
as shown below. 

Before going to iterative calculations, we mention a relation of Eq. (\ref{lenseq}) 
to an improved version of the gravitational lens equation by Bozza 
\cite{Bozza2008}. 
Eq. (\ref{lenseq}) is rearranged as 
\begin{align}
\alpha_G - \theta + B = 
\arcsin\left(\frac{D_L}{\sqrt{(D_{LS})^2 + (D_S)^2 \tan^2 \beta}} \sin \theta \right) ,
\label{lenseq2}
\end{align}
where we used Eq. (\ref{B2}). 
By taking the sine of the both sides of Eq. (\ref{lenseq2}), 
we obtain 
\begin{align}
& \sin B \cos(\alpha_G - \theta) + \cos B \sin(\alpha_G - \theta) 
\notag\\
&= 
\frac{D_L}{\sqrt{(D_{LS})^2 + (D_S)^2 \tan^2 \beta}} \sin \theta . 
\label{lenseq3}
\end{align}

By using Eqs. (\ref{base}) and (\ref{LS}) for the triangle in Figure \ref{fig-B}, 
we obtain 
\begin{align}
\sin B &= 
\frac{D_S \tan\beta}{\sqrt{(D_{LS})^2 + (D_S)^2 \tan^2 \beta}} , 
\label{sinB}
\\
\cos B&= 
\frac{D_{LS}}{\sqrt{(D_{LS})^2 + (D_S)^2 \tan^2 \beta}} . 
\label{cosB}
\end{align}
 Substituting Eqs. (\ref{sinB}) and (\ref{cosB}) into Eq. (\ref{lenseq3}) 
 leads to 
 \begin{align}
D_S \tan\beta \cos(\alpha_G - \theta) + D_{LS} \sin(\alpha_G - \theta) 
= 
D_L \sin \theta . 
\label{lenseq4}
\end{align}
It is straightforward to rearrange Eq. (\ref{lenseq4}) as 
\begin{align}
D_S \tan\beta 
= 
\frac{D_L \sin \theta - D_{LS} \sin(\alpha_G - \theta))}{\cos(\alpha_G - \theta)} . 
\label{lenseq-Bozza}
\end{align}
This is the improved expression of the lens equation by Bozza \cite{Bozza2008}. 
We should note that this expression is highly nonlinear in $\alpha_G$. 
It seems that it is not suitable for iterative calculations in terms of 
a complicated form of $\alpha_G$, e.g. in modified gravity theories 
\cite{Takizawa2020}.

Before closing this section, we mention the strong deflection, for which 
the light ray can have the winding number $N \geq 1$. 
The photon trajectory is described by the orbit equation 
$(du/d\phi)^2 = F(u)$, where $u =1/r$. 
See e.g. Eq. (26) in Reference \cite{Ishihara2016} for this equation. 
The orbit equation gives 
the angle separation from the source to the receiver as 
\begin{align}
\phi_{RS} = \pm \int_S^R \frac{du}{\sqrt{F(u)}} , 
\label{phiRS}
\end{align}
where $\pm$ correspond to the anticlockwise or clockwise motion, respectively. 
Therefore, $\phi_{RS}$ can be larger than $2\pi$. 
As a result,  also $\alpha_I$ in Eq. (\ref{alphaI}) can. 
Corresponding to this, $\alpha_G$ has a modulo $2\pi$, so that 
it can describe also the strong deflection case. 
See Reference \cite{Ishihara2017} for the strong deflection in finite-distance cases. 
For the strong deflection case, 
it should be noted that $\arcsin$ functions have a modulo $2\pi$. 
Therefore, Eq. (\ref{lenseq}) is modified for the strong deflection case as 
\begin{align}
&\alpha_G - \theta 
- \arcsin\left(\frac{D_L}{\sqrt{(D_{LS})^2 + (D_S)^2 \tan^2 \beta}} \sin \theta \right) 
\notag\\
&+ \arctan \left( \frac{D_S}{D_{LS}} \tan\beta \right)  
\notag\\
&= 2 n \pi , 
\label{lenseq-strong} 
\end{align}
where $n$ is an integer. 
Here, the sign of $n$ is chosen as the same as that of $\alpha_G$, 
such that $n$ can mean the winding number $N$ of the light ray. 
Dabrowski and Schunck mentioned also the large deflection case, 
where they assume that the intersection point Q is in the lens plane 
\cite{DS}. 
Even if the intersection point Q is in the lens plane and the source and
observer are at infinity, there are differences between their equation 
and the present results of Eqs.  (\ref{lenseq}) and (\ref{lenseq-strong}).  
This is because Dabrowski and Schunck equation is expressed 
in terms of distances between objects 
instead of distances between planes.

\section{Iterative solutions of the gravitational lens equation with finite-distance effects} 
\subsection{Iterative method for the finite-distance gravitational lens equation}
Eq. (\ref{lenseq}) is the finite-distance gravitational lens equation 
that holds for a general situation 
in a static and spherically symmetric spacetime. 
In this section, we shall examine an iterative method 
for Eq. (\ref{lenseq}). 
For this purpose, we make an additional assumption that 
all the angles of $\beta$, $\theta$ and $\alpha_G$ are small, 
namely $|\beta| \ll 1$, $|\theta| \ll 1$ and $|\alpha_G| \ll 1$.  
Note that $\alpha_G$ for a strong deflection case 
can be larger than the order of unity, 
even if $\beta$ and $\theta$ are small. 

It is convenient to introduce a (nondimensional) bookkeeping parameter $\varepsilon$ 
in order to make the present iterative procedure more transparent. 
The intrinsic source position $\beta$ is given. 
Therefore, we do not expand $\beta$ in $\varepsilon$. 
In the small angle approximation, $\beta$ is small. 
Hence, it can be expressed in terms of $\varepsilon$ as 
\begin{align}
\beta = \varepsilon \beta_{(1)} . 
\label{beta-exp}
\end{align}
On the other hand, $\theta$ and $\alpha_G$ are nonlinearly dependent 
on the intrinsic source position. 
In the small-angle approximation, hence, 
they can be expressed in a Taylor series as 
\begin{align}
\theta 
&= \sum_{k=1}^{\infty} \varepsilon^k \theta_{(k)} , 
\label{theta-exp}
\\
\alpha_G 
&= \sum_{k=1}^{\infty} \varepsilon^k \alpha_{G (k)} . 
\label{alpha-exp}
\end{align}

Eqs. (\ref{beta-exp}), (\ref{theta-exp}) and (\ref{alpha-exp}) 
are substituted into Eq. (\ref{lenseq}). 
At the first order in $\varepsilon$, we obtain the {\it linearized} 
lens equation 
\begin{align}
\beta_{(1)} = \theta_{(1)} - \frac{D_{LS}}{D_S} \alpha_{G(1)} . 
\label{lenseq-first}
\end{align}
 It seems that Eq. (\ref{lenseq-first}) is the same 
as the conventional lens equation. 
However, we should note that $\alpha_{G(1)}$ 
in Eq. (\ref{lenseq-first}) contain effects of finite distances. 
For a given $\beta$, Eq. (\ref{lenseq-first}) is an equation 
for the unknown variable $\theta_{(1)}$. 

At the second order in $\varepsilon$, 
Eq. (\ref{lenseq}) becomes a linear equation for $\theta_{(2)}$ and 
it is immediately solved as 
\begin{align}
\theta_{(2)} = \frac{D_{LS}}{D_S} \alpha_{G(2)} . 
\label{lenseq-second}
\end{align}
We thus obtain the second-order solution $\theta_{(2)}$, 
because $\alpha_{G(2)}$ is calculated by using $\theta_{(1)}$. 

At the third order in $\varepsilon$, 
Eq. (\ref{lenseq}) gives a solution for $\theta_{(3)}$. 
\begin{align}
\theta_{(3)} 
= & 
\frac{D_{LS}}{D_S} \alpha_{G(3)} 
\notag\\ 
&
+ \frac13 \left[ 1 - \left( \frac{D_S}{D_{LS}} \right)^2 \right] (\beta_{(1)})^3 
+ \frac12 \frac{D_L D_S}{(D_{LS})^2} (\beta_{(1)})^2 \theta_{(1)} 
\notag\\
&
+ \frac16 \frac{D_L}{D_S} 
\left[ 1 - \left( \frac{D_L}{D_{LS}} \right)^2 \right] 
(\theta_{(1)})^3 .
\label{lenseq-third}
\end{align}

\subsection{Einstein ring in an iterative scheme} 
The finite-distance effects on the deflection angle of light are 
discussed in e.g. References \cite{Ishihara2016, Takizawa2020}. 
In their iterative calculations for Schwarzschild, Kottler or Weyl gravity models, 
the deflection angle at the lowest order is $4m/b$, 
where $m$ is the lens mass. 
Eq. (\ref{b-theta}) is rewritten as 
$b = D_L \sin\theta = \varepsilon D_L \theta_{(1)} + O(\varepsilon^2)$. 
Eq. (\ref{alpha-exp}) in the small angle approximation 
means $\alpha_G = O(\varepsilon)$. 
Therefore, the scaling of the lens mass should be $m = \varepsilon^2 M$, 
where $M \equiv m_{(2)}$ and $M$ is independent of $\varepsilon$. 
By substituting $b = D_L \sin\theta$  and $m = \varepsilon^2 M$ 
into the form of $4m/b$, we obtain 
the linear order of $\alpha_G$ in $\varepsilon$ as 
\begin{align}
\alpha_{G(1)} = \frac{4M}{D_L \theta_{(1)}} , 
\label{alpha-lowest}
\end{align}
where we use $\theta_{(1)} \neq 0$.

Only in this paragraph, 
we assume that the source is located exactly behind the lens. 
Namely, $\beta = 0$ is assumed. 
By substituting Eq. (\ref{alpha-lowest}) into Eq. (\ref{lenseq-first}), 
we obtain a quadratic equation as 
\begin{align}
(\theta_{(1)})^2 = \frac{4 M D_{LS}}{D_L D_S} . 
\label{theta-first-2}
\end{align}
This means that the lensed image becomes a circle 
that is usually called the Einstein ring. 
Therefore, we define the radius of the Einstein ring by 
\begin{align}
\theta_{E (1)} \equiv  
\sqrt{\frac{4 M D_{LS}}{D_L D_S}} . 
\label{Einsteinring}
\end{align}
This definition is consistent with that in the conventional gravitational lens formulation 
that assumes the small angle approximation and the asymptotic receiver and source. 

Rigorously speaking, 
the expression for the Einstein ring radius for exotic objects 
such as a wormhole \cite{Abe, Toki} 
may be different from Eq. (\ref{Einsteinring}) 
for the Schwarzschild spacetime or a spacetime model that 
approaches the Schwarzschild spacetime in a certain limit. 
See e.g. also Eqs. (5) and (6) in Reference \cite{Izumi2013}, 
in which Izumi et al. discussed the radius of the Einstein ring 
for an inverse power model 
(proposed by Kitamura et al. \cite{Kitamura2013}) 
representing the Ellis wormhole 
and Schwarzschild black holes in the weak field approximation. 

We should note that the Einstein ring radius by Eq. (\ref{Einsteinring}) 
is valid only at the lowest order in iterative calculations. 
The actual radius of the Einstein ring is {\it dressed} 
in the present iteration scheme, 
because it is the sum of all the terms in $\varepsilon$, namely 
\begin{align}
\theta_{E} = \sum_{k=1}^{\infty} 
\varepsilon^k \theta_{E (k)} . 
\label{Einsteinring-dressed}
\end{align}

The discussion and expressions in this section are general. 
In the Weyl conformal gravity case, 
$\theta_{(2)}$ vanishes as shown in the next section.

\section{Weyl conformal gravity on the lens equation}
\subsection{Deflection of light in Weyl conformal gravity} 
The Weyl conformal gravity model was proposed by Bach \cite{Bach}. 
The action in the Weyl conformal gravity is written as 
\begin{align}
S = \int d^4x \sqrt{-g} C_{abcd} C^{abcd} ,
\label{Weyl-action}
\end{align}
where $g$ denotes the determinant of the metric. 
Birkoff's theorem still holds even for a generalized solution 
in the Weyl conformal gravity \cite{Riegert}. 
The static and spherically symmetric vacuum solution 
in the Weyl conformal gravity was obtained by Mannheim and Kazanas 
\cite{MK}. 
This solution is expressed by using three new parameters 
(often denoted as $\beta$, $\gamma$ and $k$). 
It is written as
\begin{align}
d s^2 = -B(r) d t^2 + B^{-1}(r) d r^2 + r^2 (d\theta^2 + \sin^2 \theta d\phi^2) . 
\label{ds-Weyl}
\end{align}
$B(r)$ can be approximated as 
\begin{align}
B(r) = 1 - 3m \gamma - \frac{2m}{r} + \gamma r -k r^2 . 
\end{align}
Here, $m \gamma \ll 1$ is assumed, so that we can neglect 
$m^2 \gamma (= m \times m\gamma \ll m)$. 
$r^2$ terms have been already discussed in the Kottler model 
\cite{Takizawa2020}. 
For the simplicity, we ignore $-k r^2$ in $B(r)$ in the following. 
Mannheim and Kazanas argued that the Weyl gravity can explain 
the flat rotation of galaxies without introducing dark matter, 
for which $\gamma$ is of the order of the inverse of the Hubble radius 
(denoted as $r_H$), namely $\gamma \sim r_H^{-1}$ \cite{MK}. 
We focus on this Weyl gravity model. 
Physically, $m \gamma \ll 1$ means that 
the black hole under study is much smaller than 
the Hubble radius of the present universe, namely 
$m \ll r_H$. 

Takizawa et al. obtained the deflection angle of light 
for the receiver and source that are within a finite distance from a lens object 
in Weyl conformal gravity \cite{Takizawa2020}. 
It is 
\begin{align} 
\alpha_{Weyl} 
=&
  \frac{2m}{b}\left(\sqrt{1 - b^2 u^2_S} + \sqrt{1 - b^2 u^2_R}\right)
\notag\\
&   - m\gamma\left(\frac{bu_S}{\sqrt{1 - b^2 u^2_S}} 
   + \frac{bu_R}{\sqrt{1 - b^2 u^2_R}}\right)
  \notag\\
&  + O\left(m^2, \gamma^2\right) , 
\label{alpha-Weyl}
\end{align}
where $u$ defines the inverse distance as $u \equiv 1/r$, 
$u_R$ and $u_S$ denote the inverse distance from the lens 
to the receiver and source, respectively. 
Note that terms linear in $\gamma$ do not exist in $\alpha_{Weyl}$. 

Several authors made attempts to calculate the deflection angle in this spacetime 
in the literature 
\cite{Edery,Pireaux2004a,Pireaux2004b,Sultana,Cattani,Ishihara2016}, 
though their discussions and methods are not self-consistent. 
For instance, they imagined the asymptotic receiver and source 
in such an asymptotically {\it nonflat} spacetime. 
In another case, only the $\phi_{RS}$ was considered.

By using Eq. (\ref{theta-exp}) for Eq. (\ref{b-theta}), 
$b$ is expanded in a series of $\varepsilon$ as 
\begin{align}
b =& D_L \sin\theta , 
\notag\\
=& 
\varepsilon D_L \theta_{(1)} 
+ \varepsilon^2 D_L \theta_{(2)} 
\notag\\
&
+ \varepsilon^3 D_L 
\left( \theta_{(3)} - \frac16 (\theta_{(1)})^3 \right) 
+ O(\varepsilon^4) . 
\label{b-exp}
\end{align}

The parameter $\gamma$ has no direct relation with $\varepsilon$. 
Consequently, $\gamma = O(\varepsilon^0)$.  
This is consistent with $m \gamma \ll 1$, 
because $m \gamma = O(\varepsilon^2)$. 

By using Eq. (\ref{b-exp}), we obtain 
\begin{align}
b u_R &= \varepsilon \theta_{(1)} + O(\varepsilon^2) , 
\label{buR}
\\
b u_S &= \varepsilon \frac{D_L}{D_{LS}} \theta_{(1)} + O(\varepsilon^3) , 
\label{buS}
\end{align}
where $D_{LS} / LS = \cos (\pi - \phi_{RS}) = 1 + O(\varepsilon^2)$ 
is used to obtain Eq. (\ref{buS}). 

By substituting Eqs. (\ref{buR}) and (\ref{buS}) 
into Eq. (\ref{alpha-Weyl}), we obtain the Taylor series of $\alpha_G$ 
in Eq. (\ref{alpha-exp}) up to the third order as 
\begin{align}
\alpha_{G (1)} 
=& 
\frac{D_S}{D_{LS}} \frac{(\theta_{E (1)})^2}{\theta_{(1)}} , 
\label{alpha-first}
\\
\alpha_{G (2)} 
=& 
- \frac{D_S}{D_{LS}} \frac{(\theta_{E (1)})^2 \theta_{(2)}}{(\theta_{(1)})^2} , 
\label{alpha-second}
\\
\alpha_{G (3)} 
=& 
- \frac{D_S}{D_{LS}} \frac{(\theta_{E (1)})^2}{\theta_{(1)}} 
\notag\\
&
\times 
\left[
\frac{\theta_{(3)}}{\theta_{(1)}} 
- \frac{(\theta_{(2)})^2}{(\theta_{(1)})^2} 
\right.
\notag\\
&~~~
\left.
+ \frac{1}{12} 
\left\{ 
1 + 3 \left( \frac{D_L}{D_{LS}} \right)^2 
+ 3 \gamma \frac{D_S}{D_{LS}} D_L  
\right\} (\theta_{(1)})^2 
\right] ,
\label{alpha-third}
\end{align}
where we use Eqs. (\ref{theta-exp}) and (\ref{Einsteinring}). 
Note the the parameter $\gamma$ appears only through 
the last term of $\alpha_{G (3)}$ in Eq. (\ref{alpha-third}).

\subsection{Lensed image positions in Weyl conformal gravity} 
By substituting Eq. (\ref{alpha-first}) into Eq. (\ref{lenseq-first}), 
we obtain 
\begin{align}
\beta_{(1)} = \theta_{(1)} - \frac{(\theta_{E (1)})^2}{\theta_{(1)}} , 
\label{lenseq-Weyl-first}
\end{align}
where $\theta_{(1)} \neq 0$. 
This equation is solved as 
\begin{align}
\theta_{(1)} = \frac12 
\left[ 
\beta_{(1)} \pm 
\sqrt{(\beta_{(1)})^2 + 4 (\theta_{E (1)})^2}
\right] .  
\label{solution-Weyl-first}
\end{align}
This is in agreement with the know results in the conventional lens theory, 
which are corresponding to the so-called primary (or plus) 
and secondary (or minus) images. 
However, Eq. (\ref{solution-Weyl-first}) is still valid also for nonasymptotic cases. 

Next, we substitute Eq. (\ref{alpha-second}) into Eq. (\ref{lenseq-second}). 
We immediately find the second-order solution as 
\begin{align}
\theta_{(2)} = 0 .  
\label{solution-Weyl-second}
\end{align}
In the following, we thus use $\theta_{(2)} = 0$. 

Finally, we substitute Eq. (\ref{alpha-third}) into Eq. (\ref{lenseq-third}) 
to obtain a linear equation for $\theta_{(3)}$. 
The solution for this equation is expressed in a long form as 
\begin{align}
\theta_{(3)} = \theta_{(3)}^S + \theta_{(3)}^W ,   
\label{solution-Weyl-third}
\end{align}
where $\theta_{(3)}^S$ and $\theta_{(3)}^W$ mean 
the third-order part only by the lens mass (without $\gamma$) 
and that by $\gamma$ parameter, respectively. 
They are  
\begin{align}
\theta_{(3)}^S =& 
\left(
1 + \frac{(\theta_{E (1)})^2}{(\theta_{(1)})^2}
\right)^{-1} 
\notag
\\
&
\times \left[
- \frac{1}{12} 
\left\{
1 + 3 \left(\frac{D_L}{D_{LS}} \right)^2 
\right\} 
(\theta_{E (1)})^2 \theta_{(1)} 
\right. 
\notag
\\
&~~~~~
+ \frac13 
\left\{
1 - \left(\frac{D_S}{D_{LS}} \right)^2 
\right\} 
(\beta_{(1)})^3 
\notag
\\
&~~~~~
+ \frac12 \frac{D_L D_S}{(D_{LS})^2} 
(\beta_{(1)})^2 \theta_{(1)} 
\notag
\\
&~~~~~
\left.
+ \frac16 \frac{D_L}{D_S} 
\left\{
1 - \left(\frac{D_L}{D_{LS}} \right)^2 
\right\} 
(\theta_{(1)})^3 
\right] ,   
\label{theta-3-S}
\\
\theta_{(3)}^W =&
3 \gamma \left(
1 + \frac{(\theta_{E (1)})^2}{(\theta_{(1)})^2}
\right)^{-1} 
\frac{D_L D_S}{D_{LS}} 
(\theta_{E (1)})^2 \theta_{(1)} . 
\label{theta-3-W} 
\end{align}

\section{Observability of the lensed image position shift due to Weyl gravity} 
The above calculations show that larger $\theta_E$ increases the third-order corrections 
including the Weyl gravity effect. 
Therefore, we consider a cluster of galaxies as a lens object 
in two cases separately. 
The first case is the so-called strong lensing, 
for which a lens system is close to the Einstein ring. 
For instance, giant arcs are observed 
near a central part of a massive cluster of galaxies. 
The second case is weak lensing, 
for which the source and the lens object are largely separated in the sky. 
This case plays a role in cosmic shear measurements. 

For the both cases, we consider a cluster of galaxies with mass 
$M \sim 10^{14} M_{\odot}$. 
For its simplicity, we assume 
$D_L \sim D_{LS} \sim 1$ Gpc, 
which means $D_S \sim 2$ Gpc. 
In the following calculations, therefore, 
$D_L/D_{LS} \sim D_S/D_{LS} \sim D_L/D_S \sim O(1)$. 
According to Eq. (\ref{Einsteinring}), 
the radius of the Einstein ring for this lens system becomes 
\begin{align}
\theta_{E (1)} \sim 10^{-4} , 
\label{Einsteinring-size}
\end{align}
which is corresponding to nearly one third arcminutes. 

The Weyl gravity model parameter $\gamma$ 
is \cite{MK} 
\begin{align}
\gamma \sim (r_H)^{-1} , 
\label{gamma}
\end{align}
where $r_H$ is the Hubble radius of the present universe, 
roughly speaking $\sim 10$ Gpc.

\subsection{Strong lensing case} 
We assume that a spherically symmetric lens system is close to the Einstein ring, 
for which Eq. (\ref{alpha-Weyl}) can be used for describing 
the gravitational deflection of light. 
The system is nearly the Einstein ring, 
such that 
we can assume 
\begin{align}
\beta_{(1)} 
&
\ll \theta_{E (1)}, 
\\
 \theta_{(1)} 
 &\sim \theta_{E (1)} .
\end{align} 
This means that 
\begin{align}
\beta_{(1)} \ll \theta_{E(1)} . 
\end{align} 
By using these conditions for Eq. (\ref{theta-3-S}), 
we obtain 
\begin{align}
\theta_{(3)}^S 
&\sim (\theta_{E (1)})^3 
\notag
\\
&\sim 10^{-12} ,
\label{theta-3-S-strong}
\end{align}
where we used Eq. (\ref{Einsteinring-size}). 
This corresponds to $O(10^{-1})$ microarcseconds. 
Hence, the third-order correction by the finite-distance effects, 
which must exist also in the theory of general relativity, 
is beyond reach of the current VLBI technology. 

Next, the third-order term purely in the Weyl conformal gravity 
is estimated as 
\begin{align}
\theta_{(3)}^W 
&\sim \gamma D_L (\theta_{E (1)})^3 
\notag
\\
&\sim 10^{-13} 
\left( \frac{D_L}{1 \mbox{Gpc}} \right) 
\left( \frac{10 \mbox{Gpc}}{r_H} \right) 
\left( \frac{\gamma}{(r_H)^{-1}} \right) 
\left( \frac{\theta_{E (1)}}{10^{-4}} \right)^3 , 
\label{theta-3-W-strong}
\end{align}
where we used Eq. (\ref{Einsteinring-size}). 
This is $\sim$ 10 picoarcseconds, 
far below the current capability of EHT ($\sim 30$ microarcseconds). 

By comparing Eqs. (\ref{theta-3-S-strong}) and (\ref{theta-3-W-strong}), 
we find the reason why $\theta_{(3)}^W$ is smaller by a factor of 10 
than $\theta_{(3)}^S$. 
$\theta_{(3)}^W$ includes an extra factor $D_L \gamma \sim D_L / r_H$, 
which is $\sim 10^{-1}$ in the above example. 
This implies that the more distant a lens system is, 
the larger the Weyl gravity effect on the gravitational lens becomes.

\subsection{Weak lensing case}
As a second example, we consider weak lensing, 
for which 
\begin{align}
\beta_{(1)} \gg \theta_{E (1)} . 
\end{align}
As an example, 
we assume $\beta_{(1)} \sim 10^{-3}$, 
which means the separation angle 
$\sim 10$ arcminutes 
for the above galaxy cluster model. 
Then, the linear-order solution by 
Eq. (\ref{solution-Weyl-first}) becomes 
\begin{align}
\theta_{(1)}^p 
&\equiv 
\frac12 
\left[ 
\beta_{(1)} +  
\sqrt{(\beta_{(1)})^2 + 4 (\theta_{E (1)})^2}
\right]
\notag
\\
 &\sim 
\beta_{(1)} , 
\label{theta-p}
\end{align}
and 
\begin{align}
\theta_{(1)}^m 
&\equiv 
\frac12 
\left[ 
\beta_{(1)} -  
\sqrt{(\beta_{(1)})^2 + 4 (\theta_{E (1)})^2}
\right]
\notag
\\
 &\sim 
- \beta_{(1)} 
\left(\frac{\theta_{E (1)}}{\beta_{(1)}} \right)^2 . 
\label{theta-m}
\end{align}
As a result, 
$|\theta_{(1)}^m| \ll \theta_{E (1)} \ll \beta_{(1)} \sim \theta_{(1)}^p$. 

By using Eqs. (\ref{theta-p}) and (\ref{theta-m}) 
in Eq. (\ref{theta-3-S}), we obtain
\begin{align}
\theta_{(3)}^{S p} 
&\sim (\beta_{(1)})^3 
\notag
\\
&\sim (\beta_{(1)})^2 \theta_{(1)}^p ,
\label{theta-3-Sp-weak}
\end{align}
and 
\begin{align}
\theta_{(3)}^{S m} 
&\sim (\beta_{(1)})^3 
\left(\frac{\theta_{E (1)}}{\beta_{(1)}} \right)^2 
\notag
\\
&\sim 
\theta_{(3)}^{S p} 
\left(\frac{\theta_{E (1)}}{\beta_{(1)}} \right)^2 .
\label{theta-3-Sm-weak}
\end{align}
By comparing Eq. (\ref{theta-3-Sp-weak}) with Eq. (\ref{theta-3-Sm-weak}), 
we find 
$\theta_{(3)}^{S m}$ is much smaller by the factor 
$(\theta_{E (1)}/\beta_{(1)})^2$  
than $\theta_{(3)}^{S p}$. 

For the above galaxy cluster model in this section, 
Eqs. (\ref{theta-3-Sp-weak}) and (\ref{theta-3-Sm-weak}) are estimated as 
\begin{align}
\theta_{(3)}^{S p} 
&\sim 10^{-9} 
\left(\frac{\beta_{(1)}}{10^{-3}} \right)^3 ,
\end{align}
which is corresponding to 
$\sim 10^2$ microarcseconds, 
and 
\begin{align}
\theta_{(3)}^{S m} 
&\sim 10^{-11} 
\left(\frac{\theta_{E (1)}}{10^{-4}} \right)^2
\left(\frac{\beta_{(1)}}{10^{-3}} \right) ,
\end{align}
which is corresponding to 
$\sim$ one microarcsecond. 
The latter value is beyond the current capability, 
while the former one is corresponding to 
$\sim 0.1$ milliarcseconds which are larger than 
the current VLBI accuracy. 
Therefore, the third-order effect as $\theta_{(3)}^{S p}$ 
can be relevant with VLBI observations. 
On the other hand, it is difficult to detect effects by $\theta_{(3)}^{S p}$ 
through weak lensing observations by optical telescopes 
that have currently the best image quality 
of $\sim 0.1$ arcseconds ($\sim 10^2$ milliarcseconds).

Next, we examine the third-order correction purely by the Weyl conformal gravity. 
Eq. (\ref{theta-3-W}) for the primary image $\theta_{(1)}^{p}$ becomes 
\begin{align}
\theta_{(3)}^{W p} 
&\sim \gamma D_L (\theta_{E (1)})^2 \beta_{(1)} 
\notag
\\
&\sim 
10^{-12} 
\left( \frac{D_L}{1 \mbox{Gpc}} \right) 
\left( \frac{10 \mbox{Gpc}}{r_H} \right) 
\left( \frac{\gamma}{(r_H)^{-1}} \right) 
\notag
\\
& ~~~~~~~~~
\times
\left( \frac{\theta_{E (1)}}{10^{-4}} \right)^2 
\left(\frac{\beta_{(1)}}{10^{-3}} \right) , 
\end{align}  
which is corresponding to $\sim 10^{-1}$ microarcseconds. 
This is far below the current EHT accuracy ($\sim 30$ microarcseconds). 
As a result, effects of the Weyl gravity model are negligible 
in the current and near-future lensing observations.

\begin{table}
\caption{
Summary of typical values for the strong and weak lensing 
by the Weyl gravity model, corresponding to Section V. 
We assume a cluster of galaxies with $\sim 10^{14}$ solar masses at $\sim 1$ Gpc.  
Two cases are considered. 
One is the strong lensing as $\beta_{(1)} \sim 0$. 
The other is the weak lensing as $\beta_{(1)} \sim 10^{-3}$. 
The angles in this table are in radians. }
\begin{center}
\begin{tabular}{l|c|c}
$\beta_{(1)}$ \:\:\:\:   & \:\:\:\: $\sim 0$ \:\:\:\: & \:\: $\sim 10^{-3}$ \:\: \\
\hline
$\theta_{E(1)}$ & $10^{-4}$ & $10^{-4}$ \\
\hline
$\theta^p_{(1)}$ & $10^{-4}$ & $10^{-3}$ \\
$|\theta^m_{(1)}|$ & $10^{-4}$ & $10^{-5}$ \\
\hline
$\theta^{Sp}_{(3)}$ & $10^{-12}$ & $10^{-9}$ \\
$|\theta^{Sm}_{(3)}|$ & $10^{-12}$ & $10^{-11}$ \\
\hline
$\theta^{Wp}_{(3)}$ & $10^{-13}$ & $10^{-12}$ \\
$|\theta^{Wm}_{(3)}|$ & $10^{-13}$ & $10^{-14}$
\end{tabular}
\end{center}
\label{table-1}
\end{table}

\section{Conclusion}
We discussed the finite-distance lens equation that is consistent with 
the deflection angle Takizawa et al. define\cite{Takizawa2020}. 
The present lens equation, 
though it is equivalent to the lens equation by Bozza \cite{Bozza2008}, 
is linear in the deflection angle 
and therefore it makes iterative calculations much simpler, 
 
As an explicit example of an asymptotically nonflat spacetime, 
we considered a static and spherically symmetric solution 
in Weyl conformal gravity. 
We focused on the Weyl gravity model relevant with 
the flat rotation of galaxies, for which  
$\gamma$ parameter in the Weyl gravity model 
is of the order of the inverse of the present Hubble radius \cite{MK}. 
For this case, we obtained iterative solutions  
for the finite-distance lens equation up to the third order. 
The effect of the Weyl gravity on the lensed image position 
begins at the third order and it is linear in the impact parameter of light. 

The deviation of the lensed image position from the general relativistic one 
is $\sim 10^{-2}$ microarcsecond for the lens and source with a separation angle of 
$\sim 1$ arcminute, 
where we consider a cluster of galaxies 
with $10^{14} M_{\odot}$ at $\sim 1$ Gpc for instance. 
The deviation becomes $\sim 10^{-1}$ microarcseconds, 
even if the separation angle is $\sim 10$ arcminutes. 
Therefore, effects of the Weyl gravity model 
are negligible in current and near-future observations of gravitational lensing. 
On the other hand, the general relativistic corrections at the third order 
$\sim 0.1$ milliarcseconds 
can be relevant with VLBI observations. 
However, the discussions in this paper 
are limited within a spherically symmetric model. 
Asymmetric cases are an open issue. 
Further study along this direction is left for future.

\begin{acknowledgments}
We are grateful to Marcus Werner for the useful discussions. 
We wish to thank Emanuel Gallo for the helpful comments on his recent works 
with his collaborators. 
We would like to thank Mareki Honma for the conversations 
on the EHT method and technology. 
We thank Yuuiti Sendouda, Ryuichi Takahashi, Masumi Kasai, Kei Yamada, 
Ryunosuke Kotaki, Masashi Shinoda, and  Hideaki Suzuki 
for the useful conversations. 
This work was supported 
in part by Japan Society for the Promotion of Science (JSPS) 
Grant-in-Aid for Scientific Research, 
No. 18J14865 (T.O.), No. 20K03963 (H.A.),  
in part by Ministry of Education, Culture, Sports, Science, and Technology,  
No. 17H06359 (H.A.)
and 
in part by JSPS research fellowship for young researchers (T.O.).  
\end{acknowledgments}

\end{document}